\documentstyle[preprint,aps]{revtex}  
\def\ut#1{\rlap{\lower1ex\hbox{$\sim$}}#1{}}  
\begin{document}  
\draft  
\title{  
\Large\bf Real sector of the nonminimally coupled  
scalar field to self-dual gravity }  
\author{
\baselineskip=15pt
Merced Montesinos$^{a,b}$\footnote{E--mail: merced@fis.cinvestav.mx}, 
Hugo A. Morales--T\'ecotl$^{c}$\footnote{
Associate member of Abdus 
Salam-ICTP, Trieste, Italy. E--mail: hugo@xanum.uam.mx},\\
Luis F. Urrutia$^{d}$\footnote{E--mail: me@aurora.nuclecu.unam.mx}
and J. David Vergara$^{d}$\footnote{E--mail: vergara@nuclecu.unam.mx}\\  
$^a$Department of Physics and Astronomy, University of 
Pittsburgh,\\ Pittsburgh, PA 15260, USA.\\
$^b$Departamento de F\'{\i}sica, Centro de Investigaci\'on y de 
Estudios Avanzados del I.P.N.,\\ 
Av. I.P.N. No. 2508, 07000 M\'exico D.F., M\'exico.\\ 
$^{c}$Departamento de F\'{\i}sica,  
Universidad Aut\'onoma Metropolitana-Iztapalapa,\\  
Apartado Postal 55-534, 09340, M\'exico D.F., M\'exico.\\  
$^{d}$Instituto de Ciencias Nucleares, Universidad Nacional  
Aut\'onoma de M\'exico, \\  
Apartado Postal  70-543, 04510,  M\'exico D.F., M\'exico.  
}  
    
\maketitle  

\begin{abstract}  
\baselineskip=15pt
A scalar field non-minimally coupled  to gravity is studied in   
the canonical framework, using self-dual  variables.   
The corresponding constraints are first class and polynomial.  
To identify the real sector of the theory, reality
conditions are implemented as second class constraints, leading to three real 
configurational degrees of freedom per space point. Nevertheless, this
realization makes non-polynomial some of the constraints. The original complex 
symplectic structure reduces to the expected real one, by using the   
appropriate 
Dirac brackets. For the sake of preserving the simplicity of the   
constraints, 
an  alternative method  preventing the use of Dirac brackets, is   
discussed. It 
consists of converting all second class constraints into first   
class by adding 
extra variables. This strategy is implemented  for the  
pure gravity case.  

\end{abstract} 
 
  
\renewcommand{\baselinestretch}{2}

\section{Introduction}  
The failure of using  perturbation theory to quantize 
General Relativity (GR) has   
led to different approaches, like string theory and  canonical 
quantization, which intend to define quantum gravity \cite{rovrev}.   
Instead of 
incorporating the remaining fundamental interactions as the former,  
the latter approach consists of just adopting standard quantum   
theory and 
classical GR.  
The spacetime structure, on the other hand, seems to be sensible to the 
non-perturbative aspects modeling it \cite{qgeo,thie}. This is in 
itself a major motivation for studying canonical GR.  
Some success has been achieved within canonical quantum gravity   
since the late 
eighties, after the introduction, by Ashtekar \cite{3Ashtekar}, of 
a set of complex canonical 
variables which greatly simplify the form of the constraints to which the 
theory is reduced. The kernel of this algebra of constraints defines
the space of physical states in the theory and   
some formal elements of it have been constructed \cite{3Ashtekar}.   
  
In spite of its simplicity, the constraints of GR in terms of Ashtekar 
variables describe  complex gravity. Ashtekar himself considered   
reducing to 
the real sector, through the introduction of an inner  
product designed to make hermitian the physical operators. However,  this 
strategy has not worked up to now, except for some particular   
cases. Other  
alternatives have been presented to avoid the use of reality   
conditions, at the 
price of a more cumbersome form of the constraints. Yet, quantum   
mechanically, 
the situation seems tractable \cite{thie}.  
  
There still remains the possibility of keeping the   
self-dual canonical formalism and trying to envisage  
how to select the real sector of the theory. Indeed this is   
possible, as it has 
been shown for pure gravity in \cite{3Hugo}, at the classical   
level. Reality 
conditions are implemented as second class constraints. The present   
work is 
devoted to show how this result is also valid in the case of a   
scalar field 
non-minimally coupled to gravity.

The strategy followed in this paper is an extension of the pure gravity 
analysis of \cite{3Hugo}, to incorporate a non-minimally coupled 
scalar field . It consists of five steps: 
(i) each complex canonical variable is splitted into one real and one   
imaginary degrees of freedom; (ii) then, every real and imaginary part is   
regarded as a 
configuration variable. An extended phase space is hence defined, where  
the  corresponding momenta will arise directly from the action.   
Notice that the 
definition of such momenta will produce  primary constraints; (iii)   
next, the 
reality conditions are implemented upon the splitting, as   
additional primary constraints. The criterion being to restrict to a real 
three-metric together with a real scalar field, during the whole   
evolution  
of the system; (iv) after imposing the conservation of the 
constraints, which amounts to obtaining all possible secondary   
constraints, a 
final classification of the full set into first and second class is   
performed. 
This may require  
the redefinition of some of them; and finally, (v) the issue of how 
to deal with the resulting set of second class 
constraints should be addressed. Since  the use of Dirac brackets   
take the 
first class constraints back to its Palatini canonical form, which is
highly non-polynomial, some alternatives should be tried.  
  
A review of how to select the real sector in phase space 
of pure self-dual gravity is made in section II. It includes 
the result of converting all second class into first class constraints, 
by introducing extra canonical variables. Thus, Dirac brackets 
are not used. Next, the 
extension of 
the method to the case of the scalar field non-minimally coupled to 
 gravity  is described in section III. 
The polynomial form of the constraints for the complex theory is 
exhibited, as well as the whole set of first and second class constraints 
describing the real sector of the theory. 
Finally,  
the last section contains some conclusions and perspectives. 
A  possibility regarding how to avoid 
the use of Dirac brackets to eliminate the second class constraints 
for the scalar field case  
is briefly discussed. A point on   
notation: $\tilde E^{ai}$ is  considered here as a complex density   
inverse 
triad of weight one, whilst $\tilde e^{ai}$ is its real part. This   
is just the 
opposite convention of ref.\cite{3Hugo}, but it has the advantage   
that adjusts 
to the rest of the literature \cite{3Ashtekar}. The number of over and
under tildes represent the positive or negative density weight ($\pm 1,
\ \pm 2$, etc.) respectively, unless it is obvious from the definition
of the different 
variables. In the case of the $\delta(\tilde x ,y)$ the tilde indicates
it is a density of weight one in the argument $x$.  
  
\section{Pure gravity}  
The Ashtekar complex canonical variables are: (i)  
$\tilde{E}^{ai}:=EE^{ai}$, with $E^{ai}$ being the triad  
($E^{ai}E^b{}_i :=q^{ab}$, $q^{ab}$ is the spatial three--metric),  
and $a,b,\dots=1,2,3$ are spatial indices, whereas $i,j,\dots=1,2,3$  
are SO(3) internal indices. Also, $E:={\rm det}{\;}E_{bj}$ with   
$E_{bj}$  
being the inverse of $E^{ai}$. (ii) $A_{ai}$ is the three--dimensional  
projection of the self-dual connection \cite{3Ashtekar},  
with associated covariant derivative ${\cal D}_a\lambda_i=  
\partial_a \lambda_i + \epsilon_{ijk} A_a{}^j \lambda^k$  
and $F_{ab}{}^i:=  
\partial_a A_{b}{}^i - \partial_b A_a{}^i + \epsilon^i{}_{jk}  
A_a{}^j A_b{}^k$ is the corresponding  
curvature.   
In terms of these variables, the self-dual action of canonical GR   
is given by  
\begin{equation}  
S = \int dtd^3x \left\{ -i\tilde{E}^{ai}\dot{A}_{ai}  
   - N {\cal S} - N^a{\cal V}_a -N^i{\cal G}_i\right\},  
\label{action}  
\end{equation}  
where  
\begin{equation}  
{\cal S}:= \epsilon_{ijk}\tilde{E}^{ai}\tilde{E}^{bj}F_{ab}{}^k,\quad  
{\cal V}_a:= \tilde{E}^b{}_j F_{ab}{}^j, \quad  
{\cal G}_i:= {\cal D}_a \tilde{E}^a{}_i\,,  
\label{const}  
\end{equation}  
are the constraints of the theory  
and $N,N^a,N^i$ are Lagrange multipliers.  
Such constraints are first class and polynomial  
in the phase space variables. Let us denote by ${\cal R}$ any 
of them and  by $\{ {\cal R}\}$ the full set.  
  
Notice that having 18 complex phase space variables $(A_{ai},   
\tilde{E}^{bj})$, 
together with 7 complex first class constraints, $\{ \cal R \}$, leaves   
us with 2 
complex configurational degrees of freedom. Then, in order to   
recover the 2 
real configurational degrees of freedom per point, further   
constraints are 
necessary. 
To this end, let us introduce the splitting  
\begin{equation}  
\tilde{E}^{ai} = \tilde{e}^{ai} + i \tilde{\cal E}^{ai}, \quad  
A_{bj} = \gamma_{bj} - i K_{bj} \,.  
\label{split}  
\end{equation}  
{}From now on, all the above thirty six  degrees of 
freedom  are taken as configuration variables in the action 
(\ref{action}). Hence, the associated canonical momenta $\Pi$  
lead to the primary constraints  
$  
\Phi_{{\cal E}ai}= \Pi_{{\cal E}ai},\;\;  
\Phi_{\gamma}{}^{ai} = \Pi_{\gamma}{}^{ai} + i \tilde e^{ai},\;\;  
\Phi_{K}{}^{ai} = \Pi_{K}{}^{ai} + \tilde e^{ai},\;\;  
\Phi_{eai} = \Pi_{eai}\, ,  
$  
which as a set is denoted by $\{ \Phi\}$. The coordinates of the total 
phase space are $Y_A=(\tilde e^{ai}, \tilde {\cal E}_{ai}, 
{\gamma}{}^{ai},  K_{ai}, \Pi_{eai},\Pi_{{\cal E}ai}, 
\Pi_{\gamma}{}^{ai},\Pi_{K}{}^{ai} )$.  
 
The reduction of the 
complex phase space  $(A_{ai}, \tilde{E}^{bj})$  to a real one is 
achieved by means of the following reality conditions  
\begin{equation}  
\psi^{ai}:=\tilde{\cal E}^{ai} = 0,\quad  
\chi_{ai}:=\gamma_{ai} - f_{ai}(\tilde e) = 0,  
\label{rc1}  
\end{equation}  
which are subsequently taken as  additional primary  constraints.  
 The constraints $\psi^{ai}$ enforces the $\tilde{E}^{ai}$ to be real,  
and hence the corresponding three--metric. The constraints   
$\chi_{ai}$ ensures that, upon evolution, $\tilde{E}^{ai}$ keeps   
being real.  
Using the compatibility  
condition between a real torsion-free connection and the triad, the  
form of $f_{ai}$ is chosen as  
$  
f_{ai}=\frac{1}{2} [ {\ut{e}}{ \ }_{ai}\ut{e}{ \   
}_c{}^j\epsilon_{jrs} -  
2 \ut{e}{ \ }_{aj}\ut{ e}{ \ }_c{}^j\epsilon_{irs} ] \tilde   
e^{dr}\partial_d 
\tilde e^{cs}.  
$  
Let us observe that $\chi_{ai}$ is not polynomial in $\tilde e^{bj}$.  
  
The full set of primary constraints is \{ $\{{\cal R}\}, \  \psi, \   
\chi, 
 \{ \Phi\}$ \}, written in terms of the real variables $Y_A$. The 
evolution of the primary constraints  does not introduce   
additional constraints. After  
redefining   
$  
\Phi_{eai}\rightarrow \Phi'{}_{eai}= \Phi_{eai}  
+ \alpha_{aibj} \Phi_{\gamma}{}^{bj} + \beta_{ai}{}{}^{bj} \chi_{bj}  
+ \eta_{aibj} \Phi_K{}^{bj},  
$  
the Poisson brackets matrix for the subset $\{\Xi\}=\{\{\Phi'\}, \   
\psi, \ \chi 
$\}   
reveals them as second class constraints. Besides  
having a simple form, 
it is a phase space independent, block diagonal matrix with non  
zero determinant.   
  
To keep  \{${\cal R}$\} as a first class set it is enough to redefine 
each element as   
$  
{\cal R}'= {\cal R} +  
\{ \Phi_{{\cal E}bj}, {\cal R} \} \psi^{bj} +  
\{ \Phi_{\gamma}^{bj}, {\cal R} \} \chi_{bj} +  
\{ \Phi'{}_{ebj}, {\cal R} \} \Phi_K{}^{bj} -  
\{ \Phi_{K}{}^{bj}, {\cal R} \} \Phi'{}_{ebj},  
$  
so that they have zero Poisson brackets with the previous second   
class subset. 
This redefinition preserves the property 
$\{ {\cal R}', {\cal Q}' \}\approx 0$, for any pair of constraints 
in $\{ {\cal R}'\}$.  
In this way, there are no additional contributions to 
the set of primary constraints $\{\Upsilon\} := \{{\cal R}'\} \cup   
\{\Xi\}$,   
which includes the reality conditions. Counting   
the independent variables   
gives 2 real configurational degrees of freedom per space point, as   
it should 
be for real GR \cite{3Hugo}.  
  
At this point Dirac's programme calls for the elimination of the second  
class constraints  
through the use of Dirac brackets. This, however, would yield a   
cumbersome  
form for the remaining constraints.  
One might avoid such treatment of the second class  
constraints by transforming them into first class constraints. To   
achieve this, 
by means of the Batalin-Tyutin  
procedure \cite{BT,das}, one adds a  new canonical pair,   
$\{Q^{ai},P_{bj}\}= \delta_a{}^b\delta_i{}^j\delta^{(3)}$, per   
original couple 
of second class   
constraints, i.e., the phase space is further enlarged with the new variables  
$\Psi_\Xi=(Q_{\cal E}{}^{ai}, Q_{\gamma ai},  Q_{ e}{}^{ai},  P_{{\cal 
E}ai},  P_{\gamma}{}^{ai},  P_{eai})$. In the present case,  
the set of first class constraints replacing  
the former second class set is  
\begin{eqnarray}  
\bar{\psi}^{ai} & := & \tilde{\cal E}^{ai} + Q_{\cal E}{}^{ai}, \nonumber\\  
\bar \Phi_{{\cal E}ai} & := & \Pi_{{\cal E}ai} - P_{{\cal E}ai},   
\nonumber\\  
\bar \chi_{ai} & := & \gamma_{ai} - f_{ai}(\tilde e) + Q_{\gamma 
ai},\nonumber\\  
\bar \Phi_{\gamma}{}^{ai} &:=& \Pi_{\gamma}{}^{ai} + i \tilde e^{ai} - 
P_{\gamma}{}^{ai}, \nonumber\\  
\bar \Phi_{K}{}^{ai} &:=& \Pi_{K}{}^{ai} + \tilde e^{ai}+ Q_{ 
e}{}^{ai},\nonumber\\  
\bar \Phi'{}_{eai} &:=& \Phi_{eai}  
+ \alpha_{aibj} \Phi_{\gamma}{}^{bj} + \beta_{ai}{}{}^{bj} \chi_{bj}  
+ \eta_{aibj} \Phi_K{}^{bj} - P_{eai} \label{barphi}, \;\;   
\end{eqnarray}  
which reduces to the original set by setting $Q^{ai}=0=P_{bj}$. Let  
us denote any  
of the constraints in (\ref{barphi}) by $\bar \Xi_\Lambda$.  Any pair 
satisfies $\left\{ \bar \Xi_\Lambda, \bar   
\Xi_{\Lambda'}\right\}=0$; i.e.   
the set (\ref{barphi}) is first class. Next, it is 
necessary to keep the set $\{{\cal R}'\}$ 
first class.  This can be done by recalling that the Poisson brackets   
matrix among 
the constraints
$\{\Xi\}$ is independent of the phase space variables  and by following 
the method of \cite{das}. Thus, one  redefines ${\cal R}'$ as   
\begin{equation}  
\bar {\cal R}' \equiv {\cal R}' (Y-\bar Y) \;\;, \label{eq:rtilde}  
\end{equation}  
where  
\begin{eqnarray}  
Y_A-\bar Y_A &:=& \left\{ \tilde e^{ai} - Q_{e}{}^{ai}, {\cal E}^{ai} 
 + Q_{\cal E}{}^{ai}, \gamma_{ai}+ Q_{\gamma ai} + Q_{e}{}^{bj}{\delta 
 f_{ai}\over \delta  \tilde e^{bj}}, K_{ai}+ P_{eai}+i   
Q_{e}{}^{bj}{\delta 
 f_{bj}\over \delta  \tilde e^{ai}}, \right. \nonumber\\  
&& \left.  \Pi_{eai} - P_{eai}-i Q_{\gamma ai}+ P_{\gamma}{}^{bj}{\delta 
 f_{bj}\over \delta  \tilde e^{ai} } +Q_{e}{}^{bj}\left({\delta^2 
 f_{ck}\over \delta  \tilde e^{ai}\delta  \tilde e^{bj} 
 }\Phi_\gamma^{ck} +i{\delta^2 
 f_{ai}\over \delta  \tilde e^{ck}\delta  \tilde e^{bj} 
 }\Phi_K^{ck}+ i{\delta 
 f_{bj}\over \delta  \tilde e^{ai} }   
\right), \right. \nonumber \\  
&& \left.  \Pi_{{\cal E}ai}-P_{{\cal E} ai},   
 \Pi_{\gamma}{}^{ai} - P_{\gamma}{}^{ai}-iQ_{e}{}^{ai}, \Pi_{K}{}^{ai} 
 \right\}\;.  
\end{eqnarray}  
The set (\ref{eq:rtilde}) is in involution with  
$\{{\bar \Xi}_{\Lambda}\}$, i.e.   
$\left\{ \bar {\cal R}', {\bar \Xi}_{\Lambda}\right\}=0$.  
Hence, the final whole set of 
constraints is first class and contains an Abelian ideal: $\{\bar   
\Xi\}$.  
The non-Abelian sector is just given by $\{ \bar{\cal R}'\}$.  
By construction, this sector preserves the   
structure of the  first class algebra among the elements of   
$\{{\cal R}'\}$. Notice that the set $\{ {\cal R}\}$ depends only on the
configurational variables $(\tilde e, {\cal E}, \gamma, K)$. In this
way, the most involved modifications to $\{ {\cal R}'\}$, via 
Eq.(\ref{eq:rtilde}), come from the terms that are proportional to
the second class constraints. It is worth emphasizing that all 
the non-polynomiality of the   
constraints $\{ 
\{\bar {\cal R}'\} \cup  \{\bar \Xi\}\}$  arises only through one function, 
which 
is $f_{ai}(\tilde e)$, appearing in the reality conditions 
(\ref{rc1}). Thus, one might think that a choice of (\ref{rc1}) in a 
polynomial 
form would solve this undesirable feature. However, as shown in 
\cite{3Hugo}, this not the case 
and one should look for alternative approaches.  

\section{Non-minimal self-dual ECKG theory}  
The second order action with  
 scalar field non--minimally  
coupled to gravity is given by   
\begin{eqnarray}  
S[ g^{ab}, \phi] = \int_M d^4 x \,\,\left\{  
 \sqrt{-g}\, {\cal R} -\frac12\,\sqrt{-g} \left(  
g^{ab} \partial_a \phi \partial_b \phi + (m^2 +\xi {\cal R} )\phi^2  
\right)  
\right\}\, ,\label{3Campo1}  
\end{eqnarray}  
where $\xi$ is a dimensionless constant. The canonical analysis of   
this action 
has been developed in  
\cite{3Kiefer}.  In a first order formalism one can adopt instead  
\begin{eqnarray}  
S[ {\omega}', e, \phi] & = & \int_M d^4 x \left\{\frac12  
e e^a_I e^b_J \Omega^2 {R'}_{ab}\,^{IJ} [{\omega}']-  
\frac12 e ( e^a_I e^{bI} \partial_a \phi\partial_b \phi +m^2 \phi^2)  
\right\}\, , \nonumber\\  
{\Omega^2} & := & 1- \xi \phi^2\,   
.\label{3Campo2}  
\end{eqnarray}  
As opposed to (\ref{3Campo1}), action (\ref{3Campo2}) gives, upon   
variation 
with respect to ${{\omega}'}_a\,^{IJ}$,   
\begin{eqnarray}  
{{\omega}'}_a\,^{IJ} & = & \omega_a\,^{IJ}(e) +K_a\,^{IJ}\, ,  
\nonumber\\  
K_a\,^{IJ} & = & \frac12 (e^I_a e^{bJ} -e^J_a e^{bI} )  
\frac{1}{\Omega^2} {\partial_b \Omega^2}\, ,\label{3Cone}  
\end{eqnarray}  
where $\omega_a\,^{IJ} (e)$ is the spin connection for pure gravity and  
$K_a\,^{IJ}$ is  the contorsion supported by the matter field   
\cite{HM}.   
To construct a first order action equivalent to the second order   
action in   
(\ref{3Campo1}), it is necessary  to add the term    
$-\frac34 \frac{1}{\Omega^2}\, e\,e^a_I e^{bI} \partial_a   
\Omega^2 \partial_b \Omega^2$ to the action (\ref{3Campo2}).  
The corresponding canonical analysis of the self--dual part of   
this modified action  
was studied in    
\cite{3Capovilla}. The conclusion in that paper, at the Hamiltonian   
level,   
is that the resulting constraints  
are non--polynomial in the phase space variables.  
In the present work, the self--dual part of the action   
(\ref{3Campo2}) is studied. Our result is that polynomial   
constraints are   
obtained, as opposed to the case in  \cite{3Capovilla}.  
  
The system to be considered is described by the  so called non-minimal  
self--dual Einstein--Cartan--Klein--Gordon (ECKG) action   
\begin{eqnarray}  
S [e, ^{+4} A, \phi] & = &  \int d ^4 x\, e\,\left\{ \, {\Omega}^2  
{^+{\Sigma}} ^{ab}\,_{IJ}  
{^{+4} F}_{ab}\,^{IJ} (^{+4} A) +  
\frac{\alpha}{2} \, \left(\,e^a_I e^{bI} \,\partial_a \phi  
\partial_b\phi +m^2 {\phi}^2\right)\right\}\, ,\nonumber\\  
{\Sigma}^{ab}\,_{IJ} & := & \frac12 (e^a_I e^b_J -e^a_J e^b_I)\,   
,\nonumber\\  
\Omega^2 & := & 1 +\alpha \xi \phi^2\, ,  
\label{MMM}  
\end{eqnarray}  
where the parameter $\alpha$ is introduced only to allow a   
rescaling of the  
scalar field $\phi$. The value $\alpha=-1$ corresponds to the   
case usually found in the literature\cite{3Kiefer,3Capovilla}.   
${^{+4} F}_{ab}\,^{IJ} (^{+4} A)$ is the curvature of the   
self--dual connection $^{+4} A$.   
  
Proceeding with the canonical analysis,  
the $3+1$ decomposition of space--time gives   
\begin{eqnarray}  
S & = & \int dt \int_{\Sigma} d^3 x\, \left\{ {\ut N} \Omega^2 \left[  
-\frac12 \epsilon^{ijk} {\widetilde E}^a_i {\widetilde E}^b_j   
\,F_{abk}   
\right]+ (-i \Omega^2  {\widetilde E}^a_i ) {\cal L}_t A^i_a \right\}   
\nonumber\\  
& & + \int dt \int_{\Sigma} d^3 x \left\{ (^{+4}A \cdot t)^i {\cal   
D}_a   
(-i \Omega^2 {\widetilde E}^a_i) +   
N^b (-i \Omega^2 {\widetilde E}^a_i ) F_{ab}\,^i\right\}\nonumber\\  
& & +  \frac{\alpha}{2}\int dt \int_{\Sigma} d^3 x\, \left\{   
{\ut N} {\widetilde E}^a_i {\widetilde E}^{bi} \partial_a \phi   
\partial_b   
\phi-\frac{1}{{\ut N}} \left[ {\cal L}_t \phi -{\cal L}_N \phi   
\right]^2 +   
{\ut N} (\sigma)^2 m^2 \phi^2  
\right\}\, .\label{Maria}  
\end{eqnarray}  
In order to get (\ref{Maria}) one performs the usual steps   
\cite{3Ashtekar}.   
${\widetilde E}^a_i$ is the densitizied inverse triad field,   
$A^i_a$ is the  
3--dimensional projection of the self--dual full connection   
(gravity +matter)  
and   
$F_{ab}\,^i$ is the corresponding curvature given by   
$F_{ab}\,^i = \partial_a  
A_b^i - \partial_b A_a^i +\epsilon^i\,_{jk} A^j_a A^k_b$.   
$(\sigma)^2$ is the  
determinant of the 3--dimensional spatial  
metric $q_{ab}$, expressed as a function of  
${\widetilde E}^a_i$ and $\epsilon^{ijk}$ is the  volume element of the  
3--dimensional internal metric $q_{ij}=\mbox{diag} (+1,+1,+1)$. Both  
sets of indices run from $1$ to $3$.  
  
{}From (\ref{Maria}) one finds the momentum variables associated   
with the  
gravitational and scalar fields. They are given by   
${\widetilde\pi}^a_i$ and  
${\widetilde\pi}_{\phi}$, where  
\begin{eqnarray}  
{\widetilde\pi}^a_i & := & -i\Omega^2 {\widetilde E}^a_i\, ,\nonumber\\  
{\widetilde\pi}_{\phi} & := & \frac{\delta S}{\delta ({\cal L}_t   
\phi)} =   
-\frac{\alpha}{{\ut N}} [{\cal L}_t \phi -{\cal L}_N \phi]\, .  
\label{Momentitos}  
\end{eqnarray}  
  
Then, plugging (\ref{Momentitos}) into (\ref{Maria}), one arrives at  
\begin{eqnarray}  
S = \int dt \int_{\Sigma} d^3 x \left\{   
{\widetilde\pi}^a_i {\cal L}_t A^i_a + {\widetilde\pi}_{\phi} {\cal   
L}_t \phi  
- \left({\ut N}^{\ast} {\widetilde{\widetilde C}} +   
(^{+4} A\cdot t)^i  {\widetilde {\cal G}}_i +N^a {\widetilde {\cal   
V}}_a\right) 
\right\}\, ,\label{3FoliaF}  
\end{eqnarray}  
where  
\begin{eqnarray}  
{\widetilde{\widetilde C}} & := & -\frac12 (\Omega^2)^2 \epsilon^{ijk}   
{\widetilde\pi }^a_i {\widetilde\pi}^b_j F_{abk}  +  
\frac{\alpha}{2} \Omega^2 {\widetilde\pi}^a_i{\widetilde\pi}^{bi}
\partial_a \phi \partial_b \phi -
\frac{1}{2\alpha} (\Omega^2)^3
({\widetilde\pi}_{\phi})^2 +  
\frac{\alpha}{2} i (\det{{\widetilde\pi}^a_i}) 
m^2 \phi^2\, ,\nonumber\\  
{\widetilde{\cal G}}_i & := & -{\cal D}_a {\widetilde\pi}^a_i\,   
,\nonumber\\  
{\widetilde {\cal V}}_a &:= & -{\widetilde\pi}^b_i F_{ba}\,^i +  
{\widetilde\pi}_{\phi} \partial_a \phi. 
\label{constcom}  
\end{eqnarray}  
are the scalar, vector and Gauss constraints, respectively.  
In (\ref{3FoliaF}), the Lagrange multiplier $\ut N$ was redefined to   
${\ut N}^{\ast} =\frac{{\ut N}}{(\Omega^2)^3}$. Note that the set of   
constraints (\ref{constcom}) is  
{\it polynomial} in the phase  
space variables $A^i_a$, ${\widetilde\pi}^a_i$, $\phi$ and  
${\widetilde\pi}_{\phi}$, where the symplectic structure is given by  
\begin{eqnarray}  
\{ A^i_a (x) ,{\widetilde\pi}^b_j (y)\} & = & \delta^b_a \delta^i_j   
\delta^3 (x,\tilde y)\, ,\nonumber\\  
\{ \phi (x), {\widetilde\pi}_{\phi} (y)\} & = & \delta^3 (x,\tilde y). 
\label{sympcom}  
\end{eqnarray}  
Let us compare the actions (\ref{action}) and (\ref{3FoliaF}). We   
find the same 
number   
of constraints  which, nevertheless, have extra terms  
containing the scalar field. Besides, the new momenta have contributions 
arising  
both from the scalar field and the gravity sector.  
  
In order to count the number of degrees of freedom of the theory  
it is necessary  to  
classify the constraints in terms of their algebra.  We find it    
convenient to use the  
smeared form of the constraints on  
$\Sigma$. If $\ut N$, $N^a$ and $v^i$ are arbitrary tensor fields,   
then one defines  
  
\begin{eqnarray}  
C(\ut{N}) & := & \int_{\Sigma} d ^3x \ut {N}\,\,  
{\widetilde{\widetilde C}}(x)\nonumber\\  
&  = & \int_{\Sigma} d ^3x \ut {N} \,\,\left\{  
 -\frac12 (\Omega^2)^2 \epsilon^{ijk} {\widetilde\pi }^a_i   
{\widetilde\pi}^b_j F_{abk}  +  
\frac{\alpha}{2} \Omega^2 {\widetilde\pi}^a_i{\widetilde\pi}^{bi} 
\partial_a \phi\partial_b \phi \right.\nonumber\\  
& & \left. \quad\quad\quad\quad\quad -  
\frac{1}{2\alpha} (\Omega^2)^3 ({\widetilde\pi}_{\phi})^2   
+\frac{\alpha}{2} i (\det{{\widetilde\pi}^a_i}) m^2 \phi^2  
\right\}\,\, ,\nonumber\\  
C(\vec N) & : = & \int_{\Sigma} d ^3 x  N^a {\widetilde{\cal V}}_a (x)  
= \int_{\Sigma} d ^3x N^a \left\{  
-{\widetilde\pi}^b_i F_{ba}\,^i +{\widetilde\pi}_{\phi} \partial_a \phi  
\right\}\,\, ,\nonumber\\  
G(v) & := & \int_{\Sigma} d ^3 x v^i   
{\widetilde{\cal G}}_i (x) =\int_{\Sigma} d ^3 x v^i  
\left\{  
-{\cal D}_a {\widetilde\pi}^a_i  
\right\}\,\, .  
\end{eqnarray}  
A combination of the vector and Gauss constraints yields the so called  
{\it diffeomorphisms} constraint. In terms of the vector field   
$\vec N$, this  
constraint has the   
form   
\begin{eqnarray}  
D(\vec N) & : = & \int_{\Sigma} d ^3 x  N^a  
\left[ {\widetilde{\cal V}}_a  + A^i_a\, {\widetilde{\cal   
G}}_i\right](x)  
= \int_{\Sigma} d ^3x N^a \left[  
-{\widetilde \pi}^b_i F_{ba}\,^{i}  
+A^i_a (-{\cal D}_b\,{\widetilde \pi}^b_i)  
\right]\, .  
\end{eqnarray}      
  
The following results are useful in dealing with the algebra of   
constraints  
\begin{eqnarray}  
\frac{\delta C(\ut{N})}{\delta A^l_c} & = & -{\epsilon}_l\,^{ij}  
{\cal D}_b ( \ut{N} (\Omega^2)^2 {\widetilde \pi}^c_i   
{\widetilde\pi}^b_j)\, ,  
\nonumber\\  
\frac{\delta C(\ut{N})}{\delta {\widetilde\pi}^c_l}  
& = & \ut{N} \left [
- (\Omega^2)^2 \epsilon ^{ljk} {\widetilde\pi}^a_j F_{ca\,k}  
+{\alpha} \Omega^2 {\widetilde \pi}^{al} \partial_a \phi \partial_c \phi
+ (\frac{\alpha}{2} i m^2 \phi^2 )
(\frac{3}{3!} {\ut{\eta}}_{\ abc} {\widetilde\pi}^a_j 
{\widetilde\pi}^b_k \epsilon^{jkl})\, ,  \right ] \nonumber\\  
\frac{\delta G(v)}{\delta A^l_c} & = & -v^i {\epsilon}_{il}\,^k   
{\widetilde\pi}^c_k\,\,,   
\quad\quad\quad \quad\quad  \quad\quad\quad\quad  
\frac{\delta G(v)}{\delta {\widetilde\pi}^c_l} = {\cal D}_c v^l\,\,\ ,  
\nonumber\\  
\frac{\delta D(\vec N)}{\delta A^l_c} & = & -{\cal L}_{\vec N}  
{\widetilde\pi}^c_l\;\;, \quad\quad\quad\quad\quad \quad   
\quad\quad\quad\quad  
\frac{\delta D(\vec N)}{\delta {\widetilde\pi}^c_l}  =   
+ {\cal L}_{\vec N} A^l_c \,\, .\label{3Varfun}  
\end{eqnarray}  
Then, by using (\ref{3Varfun}), the algebra of constraints turns   
out to be 
\begin{eqnarray}  
\left\{  C(\ut{N}) \,\, , \,\, C(\ut{M})\right\}  
& = & -D(\vec K) + G (K^a A_a)\,\, ,\nonumber\\  
\left\{  C (\ut{N} ) \,\, , \,\, D(\vec M)\right\}  
& = & - C({\cal L}_{\vec M} \ut{N} )\,\, ,\nonumber\\  
\left\{  C(\ut{N}) \,\, , \,\, G(v)\right\}  
& = & 0\,\, ,\nonumber\\  
\left\{  D(\vec N) \,\, , \,\, D(\vec M )\right\}  
& = & D ( [\vec N\, , \, \vec M])\,\, ,\nonumber\\  
\left\{  D(\vec N ) \,\, , \,\, G (v)\right\}  
& = &  G ({\cal L}_{\vec N} v)\,\, ,\nonumber\\  
\left\{  G(w) \,\, , \,\, G(v)\right\}  
& = & D ( -[w\, , \,  v])\,\, .\label{3algebra}  
\end{eqnarray}  
In (\ref{3algebra}), the vector field $\vec K$ is defined by  
$ K^a := (\Omega^2)^4  
({\widetilde\pi}^a_i {\widetilde\pi}^{bi} )  
(\ut{N} \partial_b \ut{M} - \ut{M}\partial_b \ut{N})$,  
while the commutator of internal vectors is  
$ [w,v]^i:= \epsilon^i\,_{jk} w^j v^k$. Also,  
the commutator of spatial vectors is defined by  
$ {[\vec N\, , \, \vec M ]}^a := {\cal L}_{\vec N} M^a $,   
as usual.  
  
The set of constraints (\ref{3algebra}) is  
{\it first class}. The counting of degrees of freedom leads to:  
$2 (9) +2 (1) - 2(7) = 6$   
phase space variables per point on $\Sigma$, which implies  
3 {\it complex} degrees of freedom: two for the gravitational field   
and one  
for the scalar field. To recover the real sector of the theory, i.e.  
three {\it real} degrees of freedom per point, one has to supply   
additional  
constraints on the phase space variables  
$A^i_a$, ${\widetilde\pi}^a_i$,  $\phi$,    
${\widetilde\pi}_{\phi}$. This is the subject of the following section.

\section{Real degrees of freedom} 
The real sector of the theory  is recovered 
by extending the corresponding steps developed for pure gravity in 
\cite{3Hugo}. To begin with, let us consider  the action  
(\ref{3FoliaF}) in the explicit form 
\begin{eqnarray} 
S = \int dt \int_{\Sigma} d^3 x \left\{ 
(-i \Omega^2 {\widetilde E}^a_i ) {\cal L}_t A^i_a +  
{\widetilde \pi}_{\phi} {\cal L}_t \phi -  
( {\ut N}^{\ast} {\widetilde{\widetilde C}} +  
(^{+4} A\cdot t )^i {\widetilde{\cal G}}_i + N^a {\widetilde {\cal   
V}}_a) 
\right\}\, , 
\end{eqnarray} 
where 
\begin{eqnarray} 
{\widetilde{\widetilde C}} & := & (\Omega^2)^2 \left( - 
\frac12 \epsilon^{ijk}\right)(-i \Omega^2 {\widetilde E}^a_i)  
(-i\Omega^2 {\widetilde E}^b_j) F_{abk} +  
\frac{\alpha}{2} \Omega^2 (-i\Omega^2 {\widetilde E}^a_i) 
(-i \Omega^2 {\widetilde E}^{bi}) \partial_a \phi  
\partial_b \phi\nonumber\\ 
& &  - \frac{1}{2 \alpha} (\Omega^2)^3 ({\widetilde\pi}_{\phi})^2 + 
\frac{\alpha}{2} i (-i \Omega^2)^3
(\det{{\widetilde E}^a_i}) m^2 \phi^2\, ,\nonumber\\ 
{\widetilde{\cal G}}_i & := & -{\cal D}_a (-i \Omega^2  
{\widetilde E}^a_i)\, ,\nonumber\\ 
{\widetilde {\cal V}}_a & := & - (-i \Omega^2 {\widetilde E}^b_i ) 
F_{ba}\,^i +{\widetilde\pi}_{\phi} \partial_a \phi\, .\label{3AccionM} 
\end{eqnarray} 
 
Step (i) in the construction consists of splitting each one of the 
fields involved in  (\ref{3AccionM}) into their real and imaginary   
parts, 
\begin{eqnarray} 
{\widetilde E}^{ai} = \widetilde{e}^{ai} +i {\widetilde{\cal   
E}}^{ai}\quad 
,\quad  
A_{ai}= M_{ai}+i V_{ai}\quad , 
\phi =\phi_1 +i\phi_2\quad , 
{\widetilde \pi_{\phi}} =\widetilde\pi_1 +i\widetilde\pi_2\, . 
\label{splitri} 
\end{eqnarray} 
Now the key point, implemented as part of  step (ii), 
is to promote each one of the real and imaginary parts 
to {\it independent}  variables, which implies that the  
enlarged phase space has 
$ 
2 [ 4\times 9 +4] = 80 
$ 
degrees of freedom per point  
on $\Sigma$. Next, one has to determine the corresponding momenta, 
which results in the following constraints,   
\begin{eqnarray} 
{\Phi_{{\cal E}}}_{ai} & := & {\Pi_{{\cal E}}}_{ai}\quad ,\quad 
{\Phi_e}_{ai} := {\Pi_e}_{ai}\quad ,\quad  
{\Phi_M}\,^{ai}:= {\Pi_M}\,^{ai}+i \Omega_1^2 \widetilde e^{ai}\quad , 
\quad  
{\Phi_V}\,^{ai}:= {\Pi_V}\,^{ai}- \Omega_1^2\widetilde e^{ai} 
\, ,\nonumber\\ 
{\Phi_{\pi_1}} & : = & \Pi_{\pi_1}\quad ,\quad 
{\Phi_{\pi_2}}  : =  \Pi_{\pi_2}\quad ,\quad 
{\Phi_{\phi_1}}  : =  \Pi_{\phi_1} -\widetilde\pi_1 \quad ,\quad 
{\Phi_{\phi_2}}  : = \Pi_{\phi_2} -i\widetilde\pi_1 \quad ,\nonumber\\ 
\label{3ConM} 
\end{eqnarray} 
denoted generically by $\Pi$. 
Note that ${\Pi_M}^{ai}$ and $\Pi_{\phi_2}$ are purely imaginary,   
{\it i.e.} 
there are 40 primary constraints arising from the definition of   
momenta.  
The corresponding 
symplectic structure is  
\begin{eqnarray} 
\{ {\widetilde{\cal E}}^{ai} (x) \, , \, {\Pi_{{\cal E}}}_{bj} (y)\}  
& := & \delta^a_b\delta^i_j \delta^3 (\tilde x,y)\quad ,\quad\,\,\,\, 
\{\widetilde e^{ai} (x) \, , \, {\Pi_e}_{bj} (y)\}  
:= \delta^a_b\delta^i_j \delta^3 (\tilde x,y)\quad ,\nonumber\\ 
\{ M_{ai} (x) \, , \, {\Pi_M}\,^{bj} (y)\}  
& := & \delta^b_a\delta^j_i \delta^3 (x,\tilde y)\quad ,\quad\, 
\{ V_{ai} (x) \, , \, {\Pi_V}\,^{bj} (y)\}  
:= \delta^b_a\delta^j_i \delta^3 (x,\tilde y)\quad ,\nonumber\\ 
\{ \phi_1 (x)\, ,\, \Pi_{\phi_1} (y)\}  
& := & \delta^3 (x,\tilde y)\quad ,\quad\quad\quad\quad 
\{ \phi_2 (x)\, ,\, \Pi_{\phi_2} (y)\}  
 :=  \delta^3 (x,\tilde y)\quad ,\nonumber\\ 
\{\widetilde \pi_1 (x)\, ,\, \Pi_{\pi_1} (y)\}  
& := & \delta^3 (\tilde x,y)\quad ,\quad\quad\quad\quad 
\{\widetilde \pi_2(x)\, ,\, \Pi_{\pi_2} (y)\}  
 :=  \delta^3 (\tilde x,y)\quad . 
\end{eqnarray} 
 
As for step (iii), the reality conditions are chosen here as a 
generalization of 
(\ref{rc1}), {\em i.e.} 
\begin{eqnarray} 
{\Psi_{{\cal E}}}\,^{ai} & := & {\widetilde{\cal E}}^{ai}\quad , 
\quad {\Psi_M}_{ai}:= M_{ai} -\Gamma_{ai} (\widetilde e) 
 +\frac12 \epsilon_{ij}\,^k \ut e_a{}^j \widetilde e^c{}_k  
\frac{1}{\Omega^2_1}\partial_c {\Omega}^2_1,  
\nonumber\\ 
\Psi_{\phi_2} & := & \phi_2\quad ,  
\quad \Psi_{\pi_2} := \widetilde\pi_2\quad , \label{3ConR} 
\end{eqnarray} 
where $\Gamma_{ai} (e)$ is the 3--dimensional spin connection. 
Again, $\Psi_{\cal E}{}^{ai}$  plays the role of enforcing  
$\widetilde{ E}^{ai}$ to be real and $\Psi_{Mai}$ keeps 
$\widetilde{ E}^{ai}$ real  
upon evolution. The term 
$\frac12 \epsilon_{ij}\,^k \ut e^j_a \widetilde e^c_k  
\frac{1}{\Omega^2_1}\partial_c {\Omega}^2_1 $ is the  
{\it real} contribution of matter to the full connection 
$A^i_a$: 
it is determined as the real term of the matter contribution 
to the spatial part  of $^+ A$, upon variation of (\ref{MMM}). 
Also,  $\Psi_{\phi_2}\;, \Psi_{\pi_2}$ constrain the scalar field,   
$\phi$,  
to the real sector. At this stage there are 60 primary constraints,   
20 of  
which arise from  the reality conditions (\ref{3ConR}).

In the next step (iv), it is necessary  to preserve upon evolution  
the full set of primary constraints  
(\ref{3ConM}) and (\ref{3ConR}). Before doing so, it is convenient to   
redefine some 
of them, ${\Phi_e}_{ai}$, 
 ${\Phi_{\phi_1}}$ and ${\Phi_{\pi_1}}$, as 
\begin{eqnarray} 
{\Phi' _e}_{ai}(x) & :=&  {\Phi_e}_{ai}(x) +{\lambda}_{aibj} (x,z)  
{\Phi_M}\,^{bj} (z) +{\varepsilon}_{ai}\,^{bj} (x,z) {\Psi_M}_{bj} (z) 
+\vartheta_{aibj}(x,z) {\Phi_V}\,^{bj} (z),\nonumber\\ 
{\Phi'_{\pi_1}} & := & \Phi_{\pi_1} -i \Psi_{\phi_2}\, ,\nonumber\\ 
\Phi' _{\phi_1} (x) & := & \Phi_{\phi_1} (x)+ {\cal A}_{ai} (x,z) 
{\Phi_M}^{ai} (z) + {\cal B}^{ai} (x,z) {\Psi_M}_{ai} (z) 
+ {\cal C}_{ai} (x,z) {\Phi_V}^{ai} (z)\, , 
\end{eqnarray} 
where 
\begin{eqnarray} 
\lambda_{aibj}(x,y) & = &  -\frac{\delta{\Psi_M}_{bj} (y)} 
{\delta \tilde e^{ai} (x)}\quad ,\quad 
{\varepsilon}_{ai}\,^{bj} (x,y) =i {\Omega^2_1}(y)  
\delta^b_a\delta^j_i\delta^3 (x,\tilde y)\, ,\nonumber\\  
\vartheta_{aibj}(x,y) & = & - i \frac{\delta{\Psi_M}_{bj} (y)} 
{\delta \tilde e^{ai} (x)}\quad ,\nonumber\\ 
{\cal A}_{ai} (x,y) & = &  
-\frac{\delta {\Psi_M}_{ai} (y)}{\delta \phi_1 (x)}\quad ,\quad  
{\cal B}^{ai} (x,y) = i  
\frac{\delta \Omega^2_1 (y)}{\delta \phi_1 (x)}\widetilde e^{ai} (y)\, , 
\nonumber\\ 
{\cal C}_{ai} (x,y)  & = &  
-i \frac{\delta {\Psi_M}_{ai} (y)}{\delta \phi_1 (x)}\quad . 
\end{eqnarray} 
Here, the Einstein summation convention for dummy indices  is   
extended to  
the continuous case  in such a way that it includes  
an implicit 3--dimensional integral for the repeated    
$z$ variable. Thus, for instance, the term 
${\lambda}_{aibj} (x,z) {\Phi_M}\,^{bj} (z)$  means 
$\int_{\Sigma} d^3 z\,\, {\lambda}_{aibj} (x,z) {\Phi_M}\,^{bj} (z)$. 
  
As the first step in classifying the constraints into first and   
second class  
one computes the Poisson brackets of the constraints 
${\Phi_{\cal E}}_{ai} (x)$, ${\Psi_{\cal E}}^{ai} (x)$,    
${\Phi_M}^{ai} (x)$,  
${\Psi_M}_{ai} (x)$, ${\Phi_{\phi_2}} (x)$, ${\Psi_{\phi_2}}(x)$,  
${\Phi_{\pi_2}}(x)$ ,  
${\Psi_{\pi_2}}(x)$, ${\Phi'_{\phi_1}} (x)$, ${\Phi_V}^{ai} (x)$, 
${\Phi'_e}_{ai}(x)$ and ${\Phi'_{\pi_1}} (x)$. In the standard 
notation, the result can be expressed as 
\begin{eqnarray} 
D= 
\left( 
\begin{array}{ccc} 
A & 0 & 0\\ 
0 & B & 0\\ 
0 & 0 & C 
\end{array} 
\right) 
\label{3Segunda} 
\end{eqnarray} 
where 
\begin{eqnarray} 
A =  
\begin{array}{ccccc} 
 & {\Phi_{\cal E}}_{bj} (y) & {\Psi_{\cal E}}^{bj} (y) &   
{\Phi_M}^{bj} (y) & 
{\Psi_M}_{bj} (y)\\ 
{\Phi_{\cal E}}_{ai} (x)  & 0 & -\delta^b_a \delta^j_i \delta^3   
(x, \tilde y) & 0 
& 0 \\ 
{\Psi_{\cal E}}^{ai} (x) & \delta^a_b \delta^i_j \delta^3 (\tilde x,y) & 0   
& 0 & 
0 \\ 
{\Phi_M}^{ai} (x) & 0 & 0 & 0 & -\delta^a_b \delta^i_j \delta^3
(\tilde x,y)\\ 
{\Psi_M}_{ai} (x) & 0 & 0 & \delta^b_a \delta^j_i \delta^3 (x,\tilde y) & 0 
\end{array} 
\, , 
\label{Amat} 
\end{eqnarray} 
 
\begin{eqnarray} 
B= 
\begin{array}{ccccc} 
 & {\Phi_{\phi_2}} (y) & {\Psi_{\phi_2}}(y) & {\Phi_{\pi_2}}(y) & 
{\Psi_{\pi_2}}(y)\\ 
{\Phi_{\phi_2}}(x) & 0 & -\delta^3 (\tilde x,y) & 0 & 0 \\ 
{\Psi_{\phi_2}}(x) & \delta^3 (x,\tilde y) & 0 & 0 & 0 \\ 
{\Phi_{\pi_2}}(x) & 0 & 0 & 0 & -\delta^3 (x,\tilde y)\\ 
{\Psi_{\pi_2}} (x) & 0 & 0 & \delta^3 (\tilde x,y)& 0 
\end{array} 
\, , 
\label{Bmat} 
\end{eqnarray} 
 
\begin{eqnarray} 
C =  
\begin{array}{ccccc} 
 & {\Phi'_{\phi_1}} (y) & {\Phi_V}^{bj} (y) & {\Phi'_e}_{bj}(y) &  
{\Phi'_{\pi_1}} (y)\\ 
{\Phi'_{\phi_1}} (x) & 0 & 
\frac{\delta \Omega^2_1 (y)}{\delta \phi_1 (x)}\widetilde e^{bj}   
(y) & 0  
& -\delta^3 (\tilde x,y) \\ 
{\Phi_V}^{ai}(x) & 
-\frac{\delta \Omega^2_1 (x)}{\delta \phi_1 (y)}\widetilde e^{ai}   
(x) & 0  
& -\Omega^2_1 (x) \delta^a_b \delta^i_j\delta^3 (\tilde x,y) & 0 \\ 
{\Phi'_e}_{ai} (x) & 0 & \Omega^2_1 (y) \delta^b_a \delta^j_i  
\delta^3 (x,\tilde y) & 0 & 0\\ 
{\Phi'_{\pi_1}} (x) & \delta^3 (x,\tilde y) & 0 & 0 & 0 
\end{array}\, . 
\label{Cmat} 
\end{eqnarray}

{}From the structure of the matrices (\ref{3Segunda}-\ref{Cmat})   
one concludes 
that the set of  
constraints arising  from the definition of the momenta, together with  
those which come from the 
reality conditions are {\it second class}. On the other hand,  
the original set of constraints  
${\widetilde{\widetilde C}}$, ${\widetilde {\cal G}}_i$,  
${\widetilde {\cal V}}_a$, in the complex theory, 
were the generators of  the gauge symmetry of the system. 
To show that this property remains so in the present real theory,   
one begins 
by redefining them  
in such a way  that they have zero Poisson brackets with the second   
class  
constraints. Then, it should be verified that this redefinition   
preserves  
the  first class character of the algebra among them. Recall that 
${\widetilde{\widetilde C}}$, ${\widetilde {\cal G}}_i$ and  
${\widetilde {\cal V}}_a$ depend only upon configuration variables  
(See (\ref{constcom}) and (\ref{splitri})). 
Denoting any of them  by $\cal R$, we find that the appropriate   
redefinition is 
 
\begin{eqnarray} 
{\cal R}' (x)& := & {\cal R} (x) + I^{bj} (x,z) {\Psi_M}_{bj} (z) + 
N_{bj} (x,z) {\Phi_V}\,^{bj} (z) 
+ 
O^{bj} (x,z) {\Phi' _e}_{bj} (z) +\nonumber\\ 
& & + 
H_{bj} (x,z) {\Psi_{{\cal E}}}\,^{bj} (z) + 
P(x,z) {\Psi_{\pi_2}}(z) + 
Q(x,z) {\Psi_{\phi_2}}(z)\nonumber\\ 
& &  + 
R(x,z) {\Phi' _{\phi_1}}(z) + 
S(x,z) {\Phi' _{\pi_1}}(z) \quad ,\label{3primera} 
\end{eqnarray} 
where 
\begin{eqnarray} 
I^{ai} (x,y) & = & \{ {\Phi_M}\,^{ai} (y), {\cal R} (x)\} 
\quad ,\quad 
H_{ai} (x,y)  =  \{ {\Phi_{{\cal E}}}_{ai} (y), {\cal R} (x)\} 
\quad ,\nonumber\\ 
N_{ai} (x,y)  & = & -\frac{1}{\Omega^2_1(y)} 
 \{ {\Phi' _e}_{ai} (y), {\cal R} (x)\} 
\quad ,\quad  
R(x,y)  = -\{ \Phi' _{\pi_1} (y)\, , \, {\cal R} (x)\}\quad ,\nonumber\\ 
\nonumber\\ 
O^{ai}(x,y) & = & \frac{1}{\Omega^2_1 (y)}\left[ 
\{ {\Phi_V}^{ai} (y)\, ,\, {\cal R} (x)\} +  
\frac{\delta \Omega^2_1 (y)}{\delta \phi_1(z)}\widetilde e^{ai}(y) 
\{ \Phi' _{\pi_1} (z) \, , \, {\cal R} (x)\}  
\right]\quad ,\nonumber\\ 
P(x,y) & = & \{ \Phi_{\pi_2} (y)\, , \, {\cal R} (x)\}\quad ,\quad 
Q(x,y) =  \{ \Phi_{\phi_2} (y)\, , \, {\cal R} (x)\}\quad ,\nonumber\\ 
S(x,y) & = & \{ \Phi' _{\phi_1} (y)\, ,\, {\cal R} (x)\} 
-\frac{1}{\Omega^2_1 (z)} 
\frac{\delta \Omega^2_1 (z)}{\delta \phi_1 (y)}\widetilde e^{bj}(z) 
\{ {\Phi' _e}_{bj} (z) \, ,\, {\cal R} (x) \}\quad . 
\end{eqnarray} 
The Poisson brackets between any  
pair ${\cal Q}'$ and ${\cal R}'$ is  
weakly zero. This can be shown by calculating  
\begin{eqnarray} 
\{ {\cal Q}'(x)\, ,\, {\cal R}' (y)\} & \approx & 
\{ {\cal Q}' (x) ,{\cal R} (y)\}\, = \nonumber\\ 
& & - \frac{1}{\Omega^2_1 (z)}\{{\Phi'_e}_{ai} (z), {\cal Q} (x)\}  
\{ {\Phi_V}^{ai} (z) , {\cal R} (y)\}\nonumber\\ 
& & + \frac{1}{\Omega^2_1 (z)}  
\{ {\Phi_V}^{ai}(z) , {\cal Q} (x)\} 
\{ {\Phi'_e}_{ai} (z) , {\cal R}(y) \}\nonumber\\ 
& & - \{ {\Phi'_{\pi_1}} (z) , {\cal Q} (x)\}  
\{ {\Phi'_{\phi_1}} (z) , {\cal R} (y)\}\nonumber\\ 
& & + \{ {\Phi'_{\phi_1}} (z) , {\cal Q} (x)\} 
\{ {\Phi'_{\pi_1}} (z) , {\cal R} (y)\}\nonumber\\ 
& & + \frac{1}{\Omega^2_1 (z) }\frac{\delta \Omega^2_1 (z )}{\delta   
\phi_1 
 ( \omega )}\widetilde e^{ai} (z ) 
\{ {\Phi'_{\pi_1}} (\omega ) , {\cal Q} (x)\} 
\{ {\Phi'_e}_{ai} (z) , {\cal R}(y) \}\nonumber\\ 
& & - \frac{1}{\Omega^2_1 (\omega) } 
\frac{\delta \Omega^2_1 (\omega )}{\delta \phi_1 ( z)}\widetilde   
e^{ai} (\omega 
)  
\{ {\Phi'_e}_{ai} (\omega ) , {\cal Q} (x)\} 
\{ {\Phi'_{\pi_1}} (z) , {\cal R}(y) \}\, . 
\end{eqnarray} 
 
The above result was obtained  by substituting (\ref{3primera})   
together with 
the fact that 
${\cal Q}(x)$, ${\cal R}(y)$,    
$\Psi_M$, $ \Psi_{\cal E}$, $\Psi_{\pi_2}$ and  
$\Psi_{\phi_2}$ are independent of the momenta. Now, by using the   
explicit 
form 
of 
${\Phi'_e}_{ai}$,  
$\Phi'_{\phi_1}$ and $\Phi'_{\pi_1}$ one finds, after a long (but   
otherwise 
direct) 
calculation 
\begin{eqnarray} 
\{ {\cal Q}'(x)\, ,\, {\cal R}' (y)\} & \approx & 
- \frac{1}{\Omega^2_1 (z)}\{{\Phi_e}_{ai} (z), {\cal Q} (x)\}  
\{ {\Phi_V}^{ai} (z) , {\cal R} (y)\}\nonumber\\ 
& & + \frac{1}{\Omega^2_1 (z)}  
\{ {\Phi_V}^{ai}(z) , {\cal Q} (x)\} 
\{ {\Phi_e}_{ai} (z) , {\cal R}(y) \}\nonumber\\ 
& & - \{ {\Phi_{\pi_1}} (z) , {\cal Q} (x)\}  
\{ {\Phi_{\phi_1}} (z) , {\cal R} (y)\}\nonumber\\ 
& & + \{ {\Phi_{\phi_1}} (z) , {\cal Q} (x)\} 
\{ {\Phi_{\pi_1}} (z) , {\cal R} (y)\}\nonumber\\ 
& & + \frac{1}{\Omega^2_1 (z) }\frac{\delta \Omega^2_1 (z )}{\delta   
\phi_1 
 ( \omega )}\widetilde e^{ai} (z ) 
\{ {\Phi_{\pi_1}} (\omega ) , {\cal Q} (x)\} 
\{ {\Phi_e}_{ai} (z) , {\cal R}(y) \}\nonumber\\ 
& & - \frac{1}{\Omega^2_1 (\omega) } 
\frac{\delta \Omega^2_1 (\omega )}{\delta \phi_1 ( z)}\widetilde   
e^{ai} (\omega 
)  
\{ {\Phi_e}_{ai} (\omega ) , {\cal Q} (x)\} 
\{ {\Phi_{\pi_1}} (z) , {\cal R}(y) \}\, . 
\label{qprp} 
\end{eqnarray}

The above expression is most suitably calculated in terms of  
the {\it original} complex phase space variables. The symplectic   
structure  
(\ref{sympcom}) yields 
\begin{equation} 
\{ A^i_a (x)\, ,\, {\widetilde E}^b_j (y)\}=\left\{ A_a{}^i (x), \frac{i 
\widetilde\pi^b{}_j(y)}{\Omega^2(y)} 
\right\}=\frac{i}{\Omega^2(y)}\delta^b_a\delta^i_j \delta^3(x,\tilde y), 
\end{equation} 
\begin{equation} 
\{ {\widetilde E}^a_i (x)\, ,\, {\widetilde\pi}_{\phi} (y)\} 
= \left\{ \frac{i\widetilde\pi^a{}_i(x)}{\Omega^2(x)},   
\widetilde\pi_{\phi} (y) 
\right\} 
= -\frac{1}{\Omega^2 (x)} {\widetilde E}^a_i (x)  
\frac{\delta \Omega^2 (x)}{\delta \phi(y)}. 
\end{equation} 
Upon substitution of these expressions in (\ref{qprp}) one obtains 
\begin{eqnarray} 
\{ {\cal Q}' (x) ,{\cal R}' (y)\} & \approx &  
\{A^i_a (z) , {\widetilde E}^b_j (\omega)\}\left( 
\frac{\delta {\cal Q} (x)}{\delta A^i_a (z)} 
\frac{\delta {\cal R} (y)}{\delta {\widetilde E}^b_i (z)} 
-  
\frac{\delta {\cal Q} (x)}{\delta {\widetilde E}^b_j (z)} 
\frac{\delta {\cal R} (y)}{\delta A^i_a(z)} 
\right)  
\nonumber\\ 
& & + \mbox{}\{\phi (z) ,{\widetilde\pi}_{\phi} (\omega )\} 
\left( 
\frac{\delta {\cal Q} (x)}{\delta \phi (z)} 
\frac{\delta {\cal R} (y)}{\delta {\widetilde\pi}_{\phi} (z)} 
-  
\frac{\delta {\cal Q} (x)}{\delta {\widetilde\pi}_{\phi} (z)} 
\frac{\delta {\cal R} (y)}{\delta \phi (z)} 
\right) \nonumber\\ 
& & + \mbox{}\{ {\widetilde E}^a_i (z) , {\widetilde \pi}_{\phi}   
(\omega )\} 
\left( 
\frac{\delta {\cal Q} (x)}{\delta {\widetilde E}^a_i (z)} 
\frac{\delta {\cal R} (y)}{\delta {{\widetilde\pi}_{\phi}}(\omega)} 
-  
\frac{\delta {\cal Q} (x)}{\delta {\widetilde\pi}_{\phi}(\omega)} 
\frac{\delta {\cal R} (y)}{\delta {\widetilde E}^a_i (z)} 
\right)\quad \nonumber\\ 
& \approx & \left\{ {\cal Q}(x), {\cal 
R}(y)\right\}_{(A,\pi),(\phi,\pi_{\phi})}\approx   0\,   
.\label{3primera2} 
\end{eqnarray} 
In the last line, the Poisson brackets are taken with respect to 
the original complex symplectic structure 
(\ref{sympcom}). 
Therefore, it has been shown that the Poisson brackets between any   
pair of  
constraints  
 ${\widetilde{\widetilde C'}}$,  
$ {\widetilde{\cal V'}}$, ${\widetilde{\cal G'}}$ are weakly zero.

Thus, the system is described by the following set of primary   
constraints:  
${\Phi_{\cal E}}_{ai} (x)$,  
${\Psi_{\cal E}}^{ai} (x)$,  ${\Phi_M}^{ai} (x)$,  
${\Psi_M}_{ai} (x)$, ${\Phi_{\phi_2}} (x)$, ${\Psi_{\phi_2}}(x)$,  
${\Phi_{\pi_2}}(x)$,  
${\Psi_{\pi_2}}(x)$, ${\Phi'_{\phi_1}} (x)$, ${\Phi_V}^{ai} (x)$, 
${\Phi'_e}_{ai}(x)$, ${\Phi'_{\pi_1}} (x)$,  $\widetilde{\widetilde   
{C'}}$, 
$\widetilde {{\cal V}'}$ and $\widetilde {{\cal G}'}$. Now, following the 
Dirac 
method,  the time conservation of the constraints is imposed using  the 
Hamiltonian density 
\begin{eqnarray} 
{\cal H} & = &  {\mu^{\cal E}}^{ai} {\Phi_{\cal E}}_{ai} + 
{\mu^{\cal E}}_{ai} {\Psi_{\cal E}}^{ai}  + {\mu^M}_{ai} {\Phi_M}^{ai}+ 
{\mu^{M}}^{ai}{\Psi_M}_{ai} + \mu^{\phi_2} \Phi_{\phi_2}+ 
\nu^{\phi_2} \Psi_{\phi_2}+\nonumber\\  
& & + \mu^{\pi_2} \Phi_{\pi_2} +  
\nu^{\pi_2} \Psi_{\pi_2}+ 
\mu^{\phi_1} \Phi'_{\phi_1} + 
{\mu^V}_{ai} {\Phi_V}^{ai} + 
{\mu^e}^{ai}{\Phi'_e}_{ai} + 
+\mu^{\pi_1} \Phi'_{\pi_1} + \nonumber\\ 
& & + 
N \widetilde{\widetilde {C'}} + 
N^a \widetilde{{\cal V}'_a}+ 
N^i \widetilde{{\cal G}'_i} \, , 
\end{eqnarray} 
where  no 3--dimensional integral is involved.  
 
{}From Eqs. (\ref{3primera}) and 
(\ref{3primera2}) one finds that the Poisson brackets between   
$\widetilde{\widetilde {C'}}$, $ \widetilde{{\cal V}'}$,  
$\widetilde{{\cal G}'}$ and  
$H=\int_{\Sigma} d^3 x {\cal H} (x)$  
are weakly zero. Finally, since the set of constraints   
${\Phi_{\cal E}}_{ai} (x)$, ${\Psi_{\cal E}}^{ai} (x)$,   
${\Phi_M}^{ai} (x)$,  
${\Psi_M}_{ai} (x)$, ${\Phi_{\phi_2}} (x)$, ${\Psi_{\phi_2}}(x)$,  
${\Phi_{\pi_2}}(x)$ ,  
${\Psi_{\pi_2}}(x)$, ${\Phi'_{\phi_1}} (x)$, ${\Phi_V}^{ai} (x)$, 
${\Phi'_e}_{ai}(x)$ and ${\Phi'_{\pi_1}} (x)$ is  second class, 
the Lagrange multipliers  
${\mu^{\cal E}}^{ai}$,  
${\mu^{\cal E}}_{ai}$, $ {\mu^M}_{ai}$,  
${\mu^M}^{ai}$, $\mu^{\phi_2}$,  
$\nu^{\phi_2}$, $\mu^{\pi_2}$,  
$\nu^{\pi_2}$, $\mu^{\phi_1}$, 
${\mu^V}_{ai}$, ${\mu^e}^{ai}$, 
$\mu^{\pi_1}$ are determined, and 
shown to be zero. In other words, there are no secondary constraints 
and the total  Hamiltonian density is given by  
\begin{equation} 
{\cal H}_{\rm Total} = \ut N \widetilde{\widetilde {C'}}+ N^a   
\widetilde{{\cal 
V}'}_a  
+ N^i\widetilde{{\cal G}'}_i \;, 
\end{equation} 
which is a combination of the first class constraints only. 
 
Now, let us count the physical degrees of freedom in terms of the real   
variables that
we have introduced.  
Recall that the enlarged phase space, with configuration variables 
(\ref{splitri}),  
has  dimension $2\times 9\times 4 + 2\times 4= 80$ per space point.   
Since there 
are $6\times 9+ 6\times 1= 60$  
second class constraints (\ref{3ConM}) and (\ref{3ConR}), the partially   
reduced phase 
space has dimension 20. 
The corresponding partially reduced symplectic structure  can be   
obtained by 
using  
Dirac brackets. To this end, the inverse of the 
second class constraints Poisson brackets matrix (\ref{3Segunda})   
is needed.  
Its calculation produces  
\begin{eqnarray} 
D^{-1} = 
\left( 
\begin{array}{ccc} 
A^{-1}& 0 & 0\\ 
0 & B^{-1} & 0\\ 
0 & 0 & C^{-1} 
\end{array} 
\right) 
\, , 
\label{Dinverse} 
\end{eqnarray} 
 
where 
 
\begin{eqnarray} 
A^{-1} = \left( 
\begin{array}{cccc} 
 0 & \delta^a_b \delta^i_j \delta^3 (\tilde x,y) & 0 
& 0 \\ 
-\delta^b_a \delta^j_i \delta^3 (x,\tilde y) & 0 & 0 & 
0 \\ 
0 & 0 & 0 & \delta^b_a \delta^j_i \delta^3 (x,\tilde y)\\ 
0 & 0 & -\delta^a_b \delta^i_j \delta^3 (\tilde x,y) & 0 
\end{array} 
\right)\, , 
\label{Ainverse} 
\end{eqnarray} 
\begin{eqnarray} 
B^{-1}=\left( 
\begin{array}{cccc} 
0 & \delta^3 (x,\tilde y) & 0 & 0 \\ 
-\delta^3 (\tilde x,y) & 0 & 0 & 0 \\ 
0 & 0 & 0 & \delta^3 (\tilde x,y)\\ 
0 & 0 & -\delta^3 (x,\tilde y)& 0 
\end{array} 
\right)\, , 
\label{Binverse} 
\end{eqnarray} 
 
\begin{eqnarray} 
C^{-1} = \left( 
\begin{array}{cccc} 
 0 & 0 & 0 & \delta^3 (x,\tilde y)\\ 
0 & 0 & \frac{1}{\Omega^2_1 (y)}\delta^a_b \delta^i_j \delta^3   
(x,\tilde y) & 0 \\ 
0 &  -\frac{1}{\Omega^2_1 (x)} \delta^b_a \delta^j_i \delta^3 (\tilde x,y)   
& 0 & 
-\frac{1}{\Omega^2_1 (x)}  
\frac{\delta \Omega^2_1 (x)}{\delta\phi_1 (y)}e^{bj} (x)\\ 
-\delta^3 (\tilde x,y) & 0 &  
\frac{1}{\Omega^2_1 (y)}  
\frac{\delta \Omega^2_1 (y)}{\delta\phi_1 (x)}e^{ai} (y) & 0 
\end{array} 
\right)\, . 
\label{Cinverse} 
\end{eqnarray} 
 
The use of (\ref{Dinverse}-\ref{Cinverse}), leads to 
\begin{eqnarray} 
{\{ f(x), g(y)\}^{\ast}} & = & + \{f(x), g(y)\} \nonumber\\ 
& &  
- \{ f(x), {\Phi_{\cal E}}_{ai} (z)\} \{ {\Psi_{\cal E}}^{ai} (z) ,   
g(y)\} 
+ \{ f(x), {\Psi_{\cal E}}^{ai} (z)\} \{ {\Phi_{\cal E}}_{ai} (z) ,   
g(y)\} 
\nonumber\\ 
& &  
- \{ f(x), {\Phi_M}^{ai} (z)\} \{ {\Psi_M}_{ai} (z) , g(y)\} 
+ \{ f(x), {\Psi_M}_{ai} (z)\} \{ {\Phi_M}^{ai} (z) , g(y)\} 
\nonumber\\ 
& &  
- \{ f(x), {\Phi_{\phi_2}}(z)\} \{ {\Psi_{\phi_2}}(z) , g(y)\} 
+ \{ f(x), {\Psi_{\phi_2}} (z)\} \{ {\Phi_{\phi_2}}(z) , g(y)\} 
\nonumber\\ 
& &  
- \{ f(x), {\Phi_{\pi_2}}(z)\} \{ {\Psi_{\pi_2}}(z) , g(y)\} 
+ \{ f(x), {\Psi_{\pi_2}} (z)\} \{ {\Phi_{\pi_2}}(z) , g(y)\} 
\nonumber\\ 
& &  
- \{ f(x), {\Phi'_{\phi_1}}(z)\} \{ {\Phi'_{\pi_1}}(z) , g(y)\} 
+ \{ f(x), {\Phi'_{\pi_1}} (z)\} \{ {\Phi'_{\phi_1}}(z) , g(y)\} 
\nonumber\\ 
& & 
- \{ f(x) , {\Phi_V}^{ai} (z)\}\frac{1}{\Omega^2_1 (z)} 
\{ {\Phi'_e}_{ai} (z), g(y)\} \nonumber\\ 
& & 
+ \{ f(x), {\Phi'_e}_{ai}(z)\} \frac{1}{\Omega^2_1 (z)} 
\{ {\Phi_V}^{ai} (z), g(y)\}\nonumber\\  
& & 
-\{ f(x) , {\Phi'_{\pi_1}} (z)\} \left[ 
\frac{1}{\Omega^2_1 (w)}  
\frac{\delta \Omega^2_1 (\omega)}{\delta \phi_1 (z)}e^{ai}(w) 
\right]\{ {\Phi'_e}_{ai} (w) , g(y)\} \nonumber\\ 
& & 
+\{ f(x) , {\Phi'_e}_{ai} (z)\} \left[ 
\frac{1}{\Omega^2_1 (z)}  
\frac{\delta \Omega^2_1 (z)}{\delta \phi_1 (\omega)}e^{ai}(z) 
\right]\{ {\Phi'_{\pi_1}} (\omega), g(y)\}\, , \label{3PDirac} 
\end{eqnarray} 
for the Dirac brackets of any two functions $f$ and $g$ on the enlarged phase 
space. 
Finally, upon partial reduction, 
the canonical variables are $V^i_a$,  
${\pi}^b_j= - \Omega^2_1 e^b_j$, $\phi_1$, $\pi_1$ and the reduced 
symplectic structure is just 
\begin{eqnarray} 
\{ V^i_a (x), {\pi}^b_j (y)\}^{\ast} & = &  
\delta^b_a \delta^i_j \delta^3 (x,\tilde y)\, , 
\nonumber\\ 
\{ \phi_1 (x) , {\pi_1} (y)\}^{\ast} & = & \delta^3 (x,\tilde y)\, . 
\end{eqnarray} 
In the same way as in the pure gravity case \cite{3Hugo}, 
the first class constraints (\ref{constcom}) 
turn out to be either purely real or purely imaginary in the above   
partially  
reduced phase space. Then, the physical phase space has dimension  
$20- 2\times 7= 6$, as expected for a real scalar field coupled to real 
gravity.

\section{Conclusions and perspectives} 
 
Previous results yielding the identification in phase space of the 
real sector of pure complex gravity \cite{3Hugo} have been successfully 
extended in this work to incorporate the case of a scalar field  
non minimally coupled to gravity,  
starting from  (complex) Ashtekar variables. This provides further 
support for the general validity of the method proposed. 
  
The procedure is as follows: the complex canonical variables are 
splitted into real 
and imaginary parts, each of which is taken as an independent new 
configuration 
variable. The corresponding momenta are subsequently defined from   
the action, 
leading to primary constraints. The real sector of the 
theory is next defined by introducing  appropriate reality   
conditions in the 
form 
of additional primary constraints. This is possible because the   
original  
phase space has been extended. The whole set of constraints is next  
classified into first and second class, after imposing 
the conservation of the primary constraints upon evolution.  
Finally, one faces the problem of how to conveniently deal with the   
resulting 
second class constraints, which include the reality conditions.  
 
The advantages of our approach are:  
i) Reality conditions are incorporated as true second class   
constraints within 
the canonical description of an extended phase space, uniquely   
associated to each physical system.  
ii) It leads to the standard  Dirac's method of 
counting the real physical degrees of   
freedom  arising from an originally complex theory.   
iii) Although we start from a pair of reality conditions (\ref{rc1}), 
only the first is truly an input, because the 
second condition appears as the consequence of demanding the 
conservation of the former upon evolution. 
 
As  opposed to \cite{3Capovilla}, we have presented here a theory for a 
scalar field non minimally coupled to gravity, 
leading to polynomial constraints, using Ashtekar variables.  
Unfortunately, the non polynomiality shows up after implementing    
the reality 
conditions, in the process of identifying the 
real sector of the theory . This happens  either for the non polynomial  
form of the 
reality conditions (\ref{3ConR}), or for their polynomial 
realization.  
Recently, however, certain non polynomial constraints have been shown  
to be tractable in  the quantum theory \cite{thie,3Barbero,3Barbero2}.  
Interestingly enough,   
in our case the whole non polynomiality is encoded in the single 
function $\Gamma_{ai}$ appearing in  (\ref{3ConR}). 
It certainly remains an open problem to determine whether 
or not the results presented here may provide a tractable alternative 
to deal with the quantum situation. 
 
The use of Dirac brackets, which is the standard way of  
eliminating the second class constraints, yields the expected real non 
polynomial form of the theory. For example, it leads to the 
Palatini canonical form  in the case
 of pure  gravity 
\cite{3Ashtekar}. 
To explore an alternative preventing the use of Dirac brackets in   
the pure 
gravity case,  
we have implemented the 
conversion of the full set of second class constraints into a first   
class 
set, 
following the method of \cite{das}. Thus, we have 
rewritten  pure  real gravity as a theory involving an alternative set
of first class constraints, which, for example, has not been previously
done starting from the Palatini formulation with second class
constraints. However, their physical meaning, together with their 
usefulness in a quantum theory still needs to be clarified. The method 
of \cite{das} works whenever the Poisson brackets matrix of the 
original second class constraints is independent of the phase space   
variables. 
This is indeed the case for pure gravity, but not for  
the scalar field non minimally coupled to gravity considered in 
this work. Hence, the application of the same strategy to the  latter 
theory  would first require an extension of the method in Ref.  \cite{das} . 
\section*{Acknowledgements} 
Partial support is acknowledged from CONACyT grant 3141P--E9608 and  
UNAM--DGAPA--IN100397. MM thanks all the 
members of the relativity group of the 
Department of Physics and Astronomy of the University of 
Pittsburgh for their warm hospitality.  MM's postdoctoral 
fellowship is funded through the CONACyT of M\'exico, fellow number 
91825.  Also MM thanks support from the {\it Sistema Nacional de 
Investigadores}.
 


\end{document}